# Enhanced 3D-printed holographic acoustic lens for aberration correction of single-element transcranial focused ultrasound

Marcelino Ferri[a], José M. Bravo[a], Juan V. Sánchez-Pérez[a], Javier Redondo[b]

a) Centro de Tecnologías Físicas. Universidad Politécnica de Valencia.

Camino de Vera S/N. 46020. Valencia. Spain

b) Instituto para la Gestión Integrada de las zonas Costeras. Universidad Politécnica de Valencia.

Carretera Nazaret-Oliva S/N. 46730. Grao de Gandia. Valencia. Spain

Corresponding author: Marcelino Ferri. mferri@fis.upv.es  0034 963877000 (75245)

0

**Abstract**

The correction of the aberration of transcranial focused ultrasounds is a relevant issue for enhancing various non-invasive medical treatments. Emission through multi-element phased arrays has been the most widely accepted method to reduce aberrations in recent years; however, a new disruptive technology, based on 3D printed acoustic lenses, has recently been proposed with lower cost and comparable accuracy. The number and size of transducers in phased array configurations was a bottleneck limiting the focusing accuracy, but once the submillimeter precision of the latest generation 3D printers has overcome this limitation, the challenge is now to improve the accuracy of the numerical simulations needed to design the lens. This study introduces and evaluates two improvements to the numerical model applied in previous works that proposed 3D lenses, which consist in the direct calculation of the phase pattern from the propagation of oscillating magnitudes in complex form, and in the introduction of the absorption phenomenon into the set of equations that describe the dynamics of the wave in both fluid and solid media. Numerical experiments are performed analysing the quality of the aberrated-corrected focus in different configurations. The results obtained show that the inclusion of absorption significantly improves focusing, especially where the thickness of the skull is more irregular.

**Introduction**

Since the first successful ablation of animal brain tissue transcranially using a single transducer by Fry and Goss (1980), the utility of focused ultrasound (FUS) applied to the brain through the intact skull has been demonstrated in many clinical implementations. Treatment of neurological infirmities such as Parkinson's disease (Magara et al 2014), Alzheimer's disease (Meng et al. 2017) or brain tumors (Coluccia et al. 2018, McDannold et al 2010) can be enhanced by this technique (Jolesz and McDannold 2014). The administration of FUS allows a transient and local opening of the blood-brain barrier (BBB) (Hynynen et al 2001, Choi et al 2007) improving the delivery of pharmacological substances such as monoclonal antibodies (Meng et al. 2017), anticancer therapeutic drugs (Kinoshita et al 2006), neurotrophic factors (Baseri et al 2012), therapeutic antibodies (Raymond et al 2008), adeno-associated virus (Alonso et al 2013, Wang et al 2015), and neural stem cells (Burgess et al 2011). Furthermore, High-Intensity Focused Ultrasound (HIFU) can thermally ablate brain tumours (Coluccia et al 2014). Brain ablation can reduce activity of hyper-excitable neurons in such cases as chronic neuropathic pain (Martin et al 2009) and essential tremors (Chang et al 2015, Elias et al 2013).

Success of these medical treatments is highly related with the precision of ultrasound focusing. Generation of precise focus remains a challenging topic due to the great aberrations induced by some physical characteristics of the skull: elastic heterogeneity, variable thickness, high ultrasound absorption and big acoustic impedance ratio relative to brain. First approaches to reduce aberration included the sonication after craniotomy (Guthkelch et al. 1991) resulting in an invasive technique with limited potential. Less aggressive techniques include the application of FUS without aberration correction, radiating from the more regular areas of the skull (Marquet et al. 2012). In last years two main lines of research can be pointed out for aberration correction, (i) the emission by multi-element phased arrays (Tanter et al. 1998, Clement et al. 2000, Hughes et al. 2017) and (ii) the modulation of the acoustic signals radiated by single element transducers (Kamimura et al. 2015). This second technique has been shown to be successful mainly in reducing the energy concentration in stationary waves and secondary foci, but does not seem so suitable for the correction of aberrations in the main focus. Therefore the line of research most followed is the first one, which consists on actively shaping the wavefront to correct bone-induced aberrations through the use of a multi-element transducer array whose phase could be adjusted individually. Adjustment of the phase must be done by inverse propagation, measured or simulated, from target to transducer. Initially phase patterns were assessed by physically placing a reference transducer inside the brain at target location (Thomas and Fink 1996, Tanter et al. 1996); several years after, the technique became absolutely non-invasive when it was proven that proper phase patterns could be obtained by numerical simulation (Sun and Hynynen 1998, Aubry et al 2003). Quality of simulations is directly related to the precise knowledge of the physical properties in each voxel of the digital brain and skull model. Experiments conducted to derive acoustic properties at each voxel from radiological variables obtained from medical imaging techniques, such as computed tomographies (CT) or magnetic resonances, are highly encouraged, but are out of the scope of this paper.

Two remarkable technological limitations are affecting the aberration correction by multi-element phased arrays, (i) the accuracy in the knowledge of the acoustic properties inferred by medical imaging and (ii) the size and number of singular elements that can be implemented in the phased arrays. To face this second limitation a new approach has been recently proposed by Ferri et al (2017) and Maimbourg et al. (2018), consisting on shaping the wavefront by means of high transparency 3D printed refractive ultrasound lenses. Acoustic field control by passive elements such as refractive lens or metamaterials has been successfully proved even at low frequency audible waves (Xie et al. 2016). The work published by Maimbourg et al. (2018) demonstrates that the spatial resolution, given by the voxel size, achieved by state of art 3D printers, improves dramatically the space resolution of phased arrays, driving to better focusing. Thus, instead of create arrays with an ever increasing number of elements, the acoustic lens approach can be considered as a promising technology by allowing submillimeter phase correction over a large surface with limited cost. Compared with the multi-element arrays technique, the use of a single element lens is a compact, affordable and low cost solution. Such an approach would further reduce the amount of equipment and skills required on site, at each therapeutic center. Manufacturing limitations of 3D lenses are negligible given their submillimeter precision; therefore the main front of the problem to face now is to increase the quality of the numerical approach. In the paper of Maimbourg et al. (2018) the authors suggest several of modifications to improve the accuracy of the method.

Following these suggestions, in this work we present two improvements of the model developed by Maimbourg et al (2018): first, by introducing the phenomenon of brain and cranial absorption into the set of equations and second, by recording the phase pattern by directly measuring it from the numerical propagation of acoustic magnitudes in complex form instead of measuring the time delay. To evaluate the focusing quality of the described method a numerical experiment has been carried out.

**Methods**

Computerized tomographies of human skulls. Physical parameters

Numerical experiments were conducted on an in-vivo computerized tomography (CT) of a human head. The in-plane spatial resolution of the slices is 0.49mm and interslice spacing of 0.63mm. For each voxel, a single value of the radiodensity in Hounsfield units is provided. To obtain a denser cubic simulation grid, a 3D linear interpolation of the parallelepipedic mesh is performed with a spatial step, $h$=0.244mm. Excitation frequency applied is 760kHz; thus desirable numerical error condition, $\lambda/h$>8 at the slowest medium, is fulfilled, considering the wave speed of the water.

In order to apply the elastodynamic damped equations in isotropic solids, a set of four independent parameters must be known at each voxel. This set may be, for instance, density in equilibrium $\rho$, p-wave speed $c$, attenuation coefficient $\alpha$, and shear modulus $G$. The two first parameters, given its accepted linear dependence with radiodensity in Hounsfield units (HU), are obtained by means of the expressions (Maimbourg et al. 2018)

$$c(x,y,z) = c_{water} + (c_{bone} - c_{water})\frac{HU(x,y,z) - HU_{\min}}{HU_{\max} - HU_{\min}} \qquad (1)$$

$$\rho(x,y,z) = \rho_{water} + (\rho_{bone} - \rho_{water})\frac{HU(x,y,z) - HU_{\min}}{HU_{\max} - HU_{\min}} \qquad (2)$$

where $\rho_{water}$ and $c_{water}$ are the density and p-wave speed of water at 21ºC, and $\rho_{bone}$ and $c_{bone}$ are respectively the density and p-wave speed of cortical bone at that temperature. Voxels displaying values beneath 0 HU or above 2400 HU were set respectively to 0 HU and 2400 HU, the expected values for water and cortical bone (Marsac et al. 2017). Densities and p-wave speed values are in accordance with Maimbourg et al. (2018), Connor et al. (2002), Connor and Hynynen (2004), and Pichardo et al. (2011). For cortical bone and for water we have considered, respectively

$$c_{bone} = 3100\,m/s,\ \rho_{bone} = 1900\,kg/m^3,\ c_{water} = 1485\,m/s,\ \rho_{water} = 10^3\,kg/m^3.$$

For the perfect development of the proposed model it would be desirable to know the four input magnitudes as point functions. However, although the dependence of radiodensity on density and p-wave velocity has been investigated, the relationship of radiodensity to attenuation coefficient and shear modulus has not been studied in depth. Therefore, we have applied a different approach for the attenuation coefficient and the shear modulus, consisting on the use of average values for each domain. For the attenuation coefficients at 760kHz we will consider $\alpha_{brain}$=2Np/m, and $\alpha_{skull}$=60Np/m. These values are similar to the smallest ones found in bibliography (Pichardo et al. (2011), Clement et al. (2004), Pinton et al. (2011)) in order to favour the extrapolation of the results. Thus, improvements found in real cases are expected to be better than that obtained in the experiment described here.

In the case of the shear modulus in solid domain, it is expected its dependence on radiodensity. We propose here to obtain it from its known relation with the p-wave modulus, $M$, and the Poisson's rate, $\nu$

$$G = \frac{M(1-2\nu)}{2(1-\nu)} \tag{3}$$

Where p-wave modulus is derived from radiodensity (Equation 7), and Poisson's rate is imposed constant for the skull with a value $\nu_{skull}$=0.316 to achieve a constant relation between p-wave and s-wave speeds, $c_p/c_s$=27/14, as proposed by Hughes et al. (2016).

Governing equations and numerical model

Governing equations implemented in the linear FDTD centered model, for both fluid and isotropic solid media, are

$$\frac{\partial \tau_{ij}}{\partial t} = (M-2G)\delta_{ij}(\nabla \vec{u}) + G\left(\frac{\partial u_i}{\partial x_j} + \frac{\partial u_j}{\partial x_i}\right) \tag{4}$$

$$\rho \frac{\partial u_i}{\partial t} + \sigma \cdot u_i = \sum_j \frac{\partial \tau_{ij}}{\partial x_j} \tag{5}$$

where $\tau_{ij}$ are the components of the stress tensor, $\delta_{ij}$ is the Kronecker delta, $u$ is the velocity, $\rho$ is the density in equilibrium, $G$ is the shear stress, $M$ is the p-wave modulus, and $\sigma$ is an artificial absorption parameter causing a perfectly exponential space dependent attenuation in an isotropic media. Dependence between the attenuation coefficient $\alpha$, in Nepers per meter, and the absorption $\sigma$, for both solid and fluid can be expressed as

$$\sigma = \rho \sqrt{\left(\omega + \frac{2c^2\alpha^2}{\omega}\right)^2 - \omega^2} \ , \qquad (6)$$

where $\omega$ is the angular frequency and c the p-wave speed at the media.

Constitutive (Equation 4) and dynamic (Equation 5) expressions applied at the whole computational domain are valid for both fluid and elastic media. At solid media -where shear modulus $G$ is not zero- axial stresses $\tau_{ii}$ may differ, and tangential stresses $\tau_{ij}$ are not null. On the other hand, at liquid media -where shear modulus $G$ is zero- the tangential stresses are also zero and the three axial stresses at each point take the same value, equivalent to the pressure at the fluid with opposite sign. The parameter $M$ of the set of equations represents the p-wave modulus at the solid but also the bulk modulus at fluid; both being obtained at any point of the skull and brain from the sound speed and the average density derived from the computerised tomographies, as follows

$$M = \rho c^2 \qquad (7)$$

Simulations are performed both in time reversal, to register the phase pattern required to generate the acoustic lens and in time forward to evaluate beam focusing and aberrations. In both cases excitation signal consists of a single frequency continuous burst enveloped by a half Hanning window during the first $n$ cycles. To facilitate the phase registration, excitation pressure is implemented in complex form

$$\tau_{11} = \tau_{22} = \tau_{33} = p_o \sin^2\left(\frac{\omega}{4n}\min\left(t, \frac{2n\pi}{\omega}\right)\right)\left(\cos(\omega t) + j\sin(\omega t)\right) \qquad (8)$$

where the length of the half Hanning defined is $n=10$ cycles, and the time step $\Delta t$ is computed based on a 0.3 Courant–Friedrichs–Lewy (CFL) condition at the cortical bone.

Phase shift registration and numerical design of corrective lens

The thickness of the corrective lens is derived from the phase pattern registered at lens position, obtained by numerical time reverse propagation of an acoustic wave emitted from the target point. At previous works (Maimbourg et al. 2018), where the performance of 3D printed lens has been evaluated, the thickness of the lens has been obtained as a function of the time delay of a pulse signal in its backpropagation from the target to each point of the lens. In this work, we apply a more accurate method based on the direct measurement of the phase at the points of the lens from the time reversal propagation of a single frequency wave emitted at the target point in complex form. The direct measurements of the phase in the domain $[0, 2\pi)$ cannot be performed without numerical manipulation, except if all oscillating variables are simulated in complex form where the phase information is included directly at any point of the mesh. Once achieved the stationary state, a few wave periods after the wave reaches the target, the numerically computed phase shift between a reference point and any other point of the computational domain remains remarkably constant. Phase is then 2D-unwrapped using the Goldstein's branch cut method (Ghiglia and Pritt, 1998) to avoid abrupt phase jumps at the center of the lens.

At a heterogeneous media, where there is not a single path between two particular emitting and receiving points, a characteristic impulse-response function can be assigned to each receiving point instead of obtaining a single time delay. Thus, the phase recording obtained by the method proposed here -as is done in typical laser holography- is a more suitable starting point for the generation of corrective lenses. In addition, the time delay method considering the absorption is not appropriated, since the final phase pattern is given by the interference of the wave at each point. This interference is dependent on both the phase and amplitude of each differential contribution, the latter being affected by absorption.

The thickness of the lens, $e(\theta, \varphi)$, associated to each measuring point is obtained by a linear interpolation from the unwrapped phase $\beta(\theta, \varphi)$, as follows

$$e(\theta, \varphi) = \frac{\beta(\theta, \varphi) - \beta o}{2\pi} \frac{\lambda_{lens} \lambda_{water}}{\lambda_{lens} - \lambda_{water}} \tag{9}$$

Where $\lambda_{lens}$, $\lambda_{water}$ are the wavelengths of the p-waves associated respectively with the lens and the surrounding media (generally water), and $\beta_o$ is an initial phase optimised to reduce the presence of abrupt phase changes in the central part of the lens. The thickness for 3D printed lens is computed after interpolating the Cartesian domain registered phases in the closest points of a mesh defined in spherical coordinates centered in the target point. If 3D lenses are implemented in simulations, their shape is again expressed in the Cartesian domain by linear interpolation.

Physical design of 3D printed lenses

The design procedure proposed is as follows: the external surface of the lens will be a perfectly spherical curve adapted to the curvature of the applied focused ultrasound transducer, and its internal surface will be rough so that its thickness -measured in radial direction- follows the registered phase pattern. The material used for ultrasonic lenses should be homogeneous, isotropic, with low absorption, and should have -in relation to water- a reasonable refractive behaviour and low reflection. Some materials as silicone into a 3D printed mould have been proposed by Maimbourg et al. (2018) for this purpose. However, we have decided the use of 3D printer plastics with acoustic specific impedance similar to water, but sound speed remarkably different from water. In practical cases the speed of sound in lenses should be about twice as fast as in water, because larger ratios would lead to thinner lenses that must be printed in 3D with higher precision, and smaller ratios would produce thicker lenses that would not be useful in practical cases. Among the great diversity of plastics available on the market, we have chosen the polylactic acid (PLA) for direct 3D printing of acoustic lenses.

The characterization of PLA has been carried out by means of propagation and resonance experiments conducted on several samples of 3D printed cylinders, and the results have been contrasted with the existing bibliography regarding physical properties of 3D printer inks (Farah et al. 2016). The considered values for the four physical magnitudes required to properly model the behaviour of a solid in the proposed model are $c_{PLA}$=2167m/s, $\rho_{PLA}$=1252kg·m$^{-3}$, $\alpha_{PLA}$=10Np/m and $\nu$=0.36. The measured quantities of these magnitudes imply a specific acoustic impedance value ($z_{PLA}$=0.2Mrayl) not very different from that of water, which favours the transmission of ultrasounds through the lens.

Despite the fact that at today's state of the art of 3D printing the process of printing moulds and filling with the desired material seems more feasible than the direct 3D printing of the lens, where some kind of anisotropy is inherent, we consider for the numerical simulations the behaviour of PLA lens as it may become a competitive option for a near future due to 3D printing is a booming technological sector. For the purposes of this study, the tolerance in the measurements of the parameters can be considered acceptable since we are going to do an exclusively numerical study. However, if the technique evaluated here were to be widely applied, more experiments would have to be carried out to accurately obtain the acoustic characteristics of 3D-printable plastics.

Simulation models

In the following, the method developed by us, which takes into account the absorption phenomenon, will be denoted as Time Reversal with Absorption (TRA). The obtained results will be compared with the method suggested by Maimbourg et al. (2018), which will be called here Time Reversal Undamped (TRU). Time forward simulations required are developed with absorption equations for both TRA and TRU methods.

Time reversal simulations for both methods are performed to register the phase pattern at the transducer position; but properly the phase pattern is registered at a Cartesian discretised spherical surface concentrical to the transducer and separated a distance equal to maximum lens thickness (Fig. 1b-f). Time forward simulations required in the experiments carried out are performed by two procedures (i) theoretical gold standard emission and (ii) complete simulation, which are exposed below.

In the **_theoretical gold standard emission_** we consider that each emitting voxel acts as a point source affected by the phase patterns registered in the time reversal simulation. These emitting voxels are at the surface where the phase pattern has been registered, i.e. separated from the transducer position a distance equal to maximum lens thickness. Amplitude is considered uniform at the emission surface.

In the **_complete simulation_** the emission is simulated from the position of the spherical transducer with uniform initial phase and amplitude, but the numerical media is modified by adding the Cartesian model of the lens placed in the correct position. In addition, a backing with conical symmetry is attached to the lens with the same acoustic properties as the lens itself. Backing is placed to prevent numerical noise in the emission.

**Numerical Experiment**

<u>Experimental design</u>

As aforementioned, the main goal of this work is to evaluate if the introduction of the absorption phenomenon in the equations for the time reversal simulation of the phase patterns (TRA model) allows to manufacture 3D lenses that are able to correct the skull aberration and to focus the beam better than the model without absorption (TRU model).

To reduce uncontrolled elements in this experiment that can mask our results and conclusions, we will develop our systematic study considering the focusing capability achieved by the ***theoretical gold standard emission*** procedure. Therefore the accuracy of the focusing of the aberrated-corrected focal spots performed by both proposed (TRA) and previous (TRU) methods will be evaluated (for the theoretical gold standard emission) in five different experimental configurations, called Point_1 to Point_5 (Fig. 1). Transducer-target distances range between 45.8mm and 107.3mm and transducer-target pairs are intentionally placed crossing irregular and non-symmetrical points of the skull to increase the aberration of the beams to be corrected. The five experimental configurations (Fig. 1) have been defined from the available CT scan. To develop our study we have simulated the emission from a single element focused transducer with a radius of curvature of 59mm and a diameter of aperture of 67mm; similar to that analysed by Maimbourg et al. (2018). A frequency of 760kHz has been selected since higher frequencies would precise longest computational times due to the spatial resolution required and would lead to an increase in absorption. Therefore, improvements associated with the simulation with dissipative equations, included in the TRA method, would be even better at higher frequencies.

Focusing capability is evaluated by means of several quantitative indicators -defined in next section- related with the position, shape, orientation, size and overlapping of the focus. To evaluate overlapping, the knowledge of a reference beam is required; therefore we will perform time forward simulations in a uniform media from a spherical source. The media consists in water without interposed skull and the source is a focused spherical transducer centered at the target point, with a radius equivalent to the distance transducer-focus and an aperture equal to lens aperture. This simulation is used to define the volume, $V_{ref}$, position, and shape of the ideal beam and as an indicative of the algorithm's numerical precision.

Additionally, we will perform in the experimental configuration of Point_4, the **_complete simulation_** of the emission with 3D interposed lenses by both TRA and TRU methods. The comparison will be made using the indicators defined above.

<u>Focusing indicators</u>

To assess the quality of each focal spot we will evaluate its (i) positional deviation, (ii) radius of gyration, (iii) orientation, (iv) volume and (v) energetic overlapping with the ideal focus. Each one of these properties is associated with a quantitative indicator defined in this section. Numerical parametrization of the beam has been performed for both -3dB and -6dB beam volumes and for the five evaluated parameters. This fact is in concordance with previous works where performance of 3D holographic ultrasonic lenses is evaluated (Maimbourg et al. 2018).

(i) Positional deviation: Position of the acoustic focus is represented by the barycentric coordinates of the module of the squared complex pressure within the -3dB (or -6dB) focal area

$$\vec{r}_F = \frac{\int\limits_{-3dB} \vec{r} \left| \overline{p}^2 \right| dV}{\int\limits_{-3dB} \left| \overline{p}^2 \right| dV} \quad \text{discretised as (for first coordinate)} \quad x_F = \frac{\sum\limits_{-3dB} x \left| \overline{p}^2 \right|}{\sum\limits_{-3dB} \left| \overline{p}^2 \right|} \qquad (10)$$

The longitudinal and transverse deviations of the focus (z, R) from the target point, $\vec{r}_o$, are obtained as

$$z = (\vec{r}_F - \vec{r}_o) \cdot \vec{u}_{TO} \ , \qquad R = \sqrt{(\vec{r}_F - \vec{r}_o)^2 - z^2} \qquad (11)$$

With $\vec{u}_{TO}$ being the unit vector in the direction and sense transducer-target

(ii) Radius of gyration: The shape of the focus is evaluated by the radius of gyration of the module of the squared complex pressure within the -3dB (or -6dB) focal area. This parameter is obtained respect to the barycentric point, $k$, and respect an axis parallel to the direction transducer target, passing through the barycentric point, $k_R$.

$$k = \sqrt{\frac{\int\limits_{-3dB} (\vec{r} - \vec{r}_F)^2 \left| \overline{p}^2 \right| dV}{\int\limits_{-3dB} \left| \overline{p}^2 \right| dV}} \qquad (12)$$

$$k_R = \sqrt{\frac{\int\limits_{-3dB}\left(\left(\vec{r}-\vec{r}_F\right)^2 - \left(\left(\vec{r}-\vec{r}_F\right)\cdot\vec{u}_{TO}\right)^2\right)\cdot\left|\overline{p}^2\right|dV}{\int\limits_{-3dB}\left|\overline{p}^2\right|dV}} \qquad (13)$$

being $\vec{r}_F$ the barycentre as defined by Equation 10 and $\vec{u}_{TO}$ the aforementioned unit vector

(iii) Orientation: To obtain the orientation, $\varphi$, we have computed, at the barycentre, the inertia tensor of the module of the squared complex pressure at the -3dB spot (or -6dB). Then the direction of the beam is defined as the direction of the eigenvector associated with the smallest eigenvalue of the tensor.

(iv) Volume: It is obtained, for both -3db and -6dB focal regions, without spatial interpolation. i.e.: voxels surrounding the target point with a squared pressure amplitude bigger than one half of the maximum (-3dB) are considered, and the rest are excluded (Fig. 2c-d).

(v) Energetic Overlapping: Denoted as $I_i(\%)$, is obtained computing the percentage of the energy of a particular beam that reaches the reference beam

$$I_i(\%) = \frac{\int\limits_{Vref}\left|\overline{p}^2\right|dV}{\int\limits_{Vi}\left|\overline{p}^2\right|dV}100 \qquad \text{discretised as} \qquad I_i(\%) = \frac{\sum\limits_{Vref}\left|\overline{p}^2\right|}{\sum\limits_{Vi}\left|\overline{p}^2\right|}100 \qquad (14)$$

Where $V_{\text{ref}}$ and $V_{\text{i}}$ are respectively the volume of the -3dB (or -6dB) focal region computed for both the reference and the aberrated-corrected scans and the squared pressure amplitude is forced to be zero out of the -3dB focal region.

**Results**

In this section we present the results of the quantitative indicators for the five configurations evaluated by ***theoretical gold standard emission*** and of the configuration Point_4 evaluated by ***complete simulation***.

<u>Theoretical gold standard emission</u>

**Positional deviation of the focus:** The longitudinal and transverse focal point deviations (z and R respectively), for the -3 dB and -6 dB beams, are shown in Tables 1 and 2 and in Figure 3a and 3b for the aberrated-corrected simulations and for the reference beam. The reference beam deviations are presented as an approximation of the numerical error of the entire time-reverse-time-forward simulation process.

The transverse deviations obtained for the -3dB beams are relatively small compared to the emitted wavelength ($\lambda_{water} \approx 2\text{mm}$ at 760kHz) and the size of the voxel in the five configurations evaluated for both methods (TRA and TRU). In all five configurations, the TRA method achieves better accuracy for transverse positioning, showing an average transverse positional deviation of 0.07mm lower than the TRU method. This value is more than twice the mean transverse deviation of the reference beam (0.03mm) which represents a rough estimate of the numerical error. In the case of -6dB focal points, no clear differences are obtained between the TRA and TRU methods.

From the results obtained, it can be seen that the measured longitudinal deviations show certain dependence on the distance from the target point. This may be related to (i) the numerical error in greater integration domains, (ii) the analytical asymmetry (bigger at larger distances) of the ideal beam in longitudinal direction, and (iii) a lower focusing capability related to the smaller solid angle of the lens subtended from the focal point. On the other hand, there are no remarkable differences between the longitudinal deviations achieved by both TRA and TRU methods.

**Radius of gyration:** The appropriate beam shape, considered as the similitude with the shape of the reference beam, is evaluated using the transverse, $k_R$, and total, $k$, radii of gyration of the -3dB and -6dB focal beams. Both the transverse and total radii of gyration at the focal point of -3 dB, simulated by the TRA and TRU methods, have certain similarities with those of the reference beam. Particularly the TRA values are generally smaller, more similar to the reference and with a more regular trend compared to TRU, as can be seen in Figure 3c for the transverse radius of gyration, and in Figure 3d for the total radius of gyration.

When -6dB focal points are evaluated, the aberrations of both methods are greater. Therefore, the trends and values of the calculated radii of gyration are in worst accordance with the ideals and, consequently, there are no clear differences between the TRA and TRU methods, as one can see in Table 2.

**Orientation:** Figure 3e shows that the -3dB beams obtained by the TRA method are better oriented than the TRU beams for the five cases evaluated. The improvements in orientation provided by the TRA method are slightly more noticeable at the outermost points. The average improvement in orientation between the TRA and TRU methods is 0.8°, with the average numerical error (reference beam orientation) being approximately 0.3°. In the -6dB beams, larger aberrations are found, with values as large as 46º for the TRA beam and 19º for the TRU beam, as shown in Table 2.

**Focal volume:** The -3dB beam focal volumes (Fig. 3f) obtained by the TRA and TRU methods show a reasonable similarity with the reference beams. When target points are close to the transducer there are no noticeable differences between the TRA and TRU methods, but at the farthest target points the TRA method provides smaller foci with a size more similar to that of the reference. The average values of the sizes calculated by the TRA method ($28mm^3$) are smaller than those obtained by the TRU method ($32mm^3$).

As with other indicators, the advantage of the TRA method over this indicator is less significant for -6dB beams than for -3dB beams, as shown in Table 2. In fact, the average volumes are $123mm^3$ for TRA and $126m\ m^3$ for TRU.

**Energetic overlapping:** In the -3dB beams (Fig. 3g) we can see that the average overlapping is 74% for the TRA method and 71% for the TRU method. In the three experimental configurations where the target point is closer to the transducer there are no notable differences between the TRA and TRU methods, but in Point_4 and Point_5 the overlaps achieved by the TRA method are significantly greater than those obtained by the TRU method. When -6 dB beams are evaluated (Table 2), the average overlaps are then 64.4% for TRA and 63.8% for TRU.

Complete simulation

In this section we present the quantitative focusing indicators of the ***complete simulation*** performed at experimental configuration Point_4 (Fig. 4c and Fig. 5c-d illustrate respectively a transversal section of the 3D computational domain and two coronal sections of the energy distribution during the complete simulation. The shape of the designed lens and the thickness distributions can be seen in Fig 4a-b). This point has been chosen because it is representative of a case in which the aberrations induced by the skull are relevant but physically resolvable, unlike other points such as Point_3 (Fig. 1d), in which there are physical limitations related to some incident angles that are impossible to be solved by a lens with the established aperture.

The focusing indicators of the aberrated-corrected focal spot are shown at Table 3. As expected, all the values are worse than those obtained by the ***theoretical gold standard emission*** procedure at this Point_4; but the TRA method still dominates the TRU method. In fact, when the - 3dB beam of ***complete simulation*** is evaluated we can see that TRA method leads to smaller positional deviation, a smaller beam volume, a more accurate orientation, a more appropriate shape (radius of gyration) and a greater energetic overlapping than TRU method. Particularly, an overlapping 20% bigger at TRA method than at TRU method must be pointed out.

**Discussion**

Numerical parameterization of the focusing has been performed for both -3dB and -6dB beams emitting a frequency of 760kHz. As expected, larger aberrations are found in the -6dB beams leading to orientations, positions and shapes of the focal point more different from the reference beam than those at -3dB. These aberrations occur regardless of the numerical method (TRA or TRU) applied to record phase patterns. On the other hand, the focusing indicators of the -6dB points calculated with the TRA method are slightly better on average than those of the TRU method, but the difference between the two methods is not as noticeable as in the -3dB beam case.

Evaluation of focusing indicators of the -3dB beam for both methods (TRA and TRU) shows several evidences to be pointed out. The most direct parameter to qualify the focusing is the energetic overlapping, and we can appreciate that in three of the five transducer-target pairs evaluated this parameter is remarkably better with the TRA method (Fig. 3g), showing its dominance over larger distances and being both methods almost equivalent for short distances. This makes sense because the aberrations created by the skull deflect the rays, but some distance is needed to obtain, from a particular angular deflection, a large transverse positional deviation. In this sense, the outermost points are of greater interest to this study and the fact that it is in the outermost points where TRA shows better behaviour than TRU is an important evidence that the TRA method is definitely better in a general comparison.

It can be also appreciated that, in the case of the TRA method, overlapping shows a rough trend to increase as the transducer-target distance increases, which may seem counterintuitive, as a larger solid angle of the transducer is known to improve focusing. But since the beam volume increases parabolically with the distance transducer-target, at longer distances we have larger focal spots, so a small positional deviation does not significantly affect the overlapping. Conversely, the same transverse deviation at closer points -where ideal beams as small as 5,239mm[3] have been found- can lead to bad overlaps.

Note that in the Point_3 we found a greater overlapping with the TRU method, but this is not related to a better focusing but to a paradoxical result in the size of the beam found, notably smaller than the reference method. In fact, the ideal beam volume is $V$=17.47 mm[3], while the TRU beam volume is $V$=9.809 mm[3]. This same particularity can be seen in the radius of gyration -both transversal and total- whose values are lower in the TRU beam than in the reference one (Table 1). Therefore, the large percentage of energy of the aberrated-corrected TRU beam contained in the reference beam is associated with the larger size of the second and not with a large similarity between the two focal spots. The transverse deviation at point_3 is smaller in TRA method ($R$=0.074mm) than in TRU ($R$=0.094mm), which reinforces the stated hypothesis that a large overlapping is not representative of a better focusing. The longitudinal deviation in this particular paradoxical case is smaller in TRU method, but as already mentioned, this parameter is not as decisive as the transverse deviation. The orientation of the TRA beam in this pair is also better than that of the TRU beam, reinforcing the stated hypothesis: In this case a better overlapping does not represent a better focusing but simply an unexpectedly small, poorly focused and poorly oriented beam. Figures 5a and 5b represent the aberrated-corrected energy distributions in Point_1 and Point_3 respectively.

The rest of the indicators evaluated can be divided into two groups depending on whether or not they show dependence on the distance transducer-target. In theory, the transverse positional deviation and orientation do not depend on the distance, whereas the radius of gyration and the volume do. On the other hand, a slight dependence on distance has been found for the longitudinal deviation (Fig. 3a). This is because the beam is not symmetrical and this asymmetry results in a longitudinal positional deviation of the barycentre depending on the size of the focus, and therefore on the distance.

The obtained results for both the transverse positional deviation and the orientation at -3dB (Fig. 3b and 3e) show that both indicators (i) have better results in TRA than in TRU in the five positions and (ii) increase roughly with distance. This increase is due to the inaccuracy of the aberration-correction, since the greater the distance to the target, the greater the angular and spatial effect of the aberration and the smaller the correction capability of the lens since the solid angle of focus is smaller. In addition, the difference between TRA and TRU methods increases roughly with distance, showing that the more critical the situation, the greater the improvement achieved by the TRA method. The statistical significance of TRA improvements over TRU for these parameters has not been studied in depth, but it should be noted that these improvements are greater than the values found for the reference beam, which are a gross reference to numerical uncertainty.

Regarding the parameters dependent on distance, volume and radius of gyration, we can see (Fig. 3) that (i) the trends of these parameters obtained by TRA are closer to the respective ideal trends than those obtained by TRU; (ii) the trends obtained by TRA are more continuous than those obtained by TRU, and (iii) the values of both indicators associated with TRA are on average smaller than those associated with TRU, indicating a greater energy dispersion in TRU, except for the paradoxical pair Point_3.

Although the number of data acquired is not sufficient for a statistically significant study of the improvement achieved by TRA with respect to TRU, we can conclude that the TRA method is generally more accurate for all indicators and at all points, except the paradoxical third point. In addition, in the present study we have attacked the skull from complicated angles and positions. In particular, the poor results obtained in Point_3 are due to the analytical impossibility of a perfect focusing, and not to inaccuracies in the model.

Regarding the ***complete simulation*** carried out at Point_4, it is worth highlighting certain aspects concerning to both the design of the 3D lens and the values of the focusing indicators obtained. The design of the 3D lens from a recorded phase pattern is an open question. Here we have defined the maximum thickness as that related to a maximum phase shift of $2\pi$, so several abrupt edges can be found. To avoid them we can use thicker lenses -which decrease transparency- or different materials with higher wave speed, with which thinner lenses could be designed that could be more affected by the spatial discretization of the mesh. In this work, no systematic lens optimization has been performed, simply a suitable common material (PLA) has been selected and its maximum thickness has been defined as indicated above. As far as the focus indicators are concerned, the overlapping achieved by the TRA method in the ***complete simulation*** (58.85%) is comparable to that obtained by the TRU method in the ***theoretical gold standard*** (63.01%), which gives the idea that the whole process of designing the lens from a registered pattern contributes to the errors in a way comparable to the error introduced by the exclusion of the absorption of the equations. This appreciation with a single indicator is merely indicative, but justifies the aim of this work to improve the physical description of the simulated phenomenon in order to enhance the medical success of the technique.

**Conclusions**

The stated objective of this study is to numerically demonstrate that the correction of skull-induced ultrasound aberrations by a specific acoustic lens adapted to a single element transducer can be significantly improved by modifying the numerical data acquisition process required for lens design. Two particular improvements are introduced: (i) the implementation of absorption in the set of equations applied for the time reversal propagation from target to transducer, and (ii) the direct recording of phase shift at each voxel of the target area, instead of the time delay recording. Our numerical experiments in five different situations taken from a human skull CT demonstrated that, compared to previously published methods, aberration correction is improved, obtaining better positioned, better oriented focal beams with a shape and size more similar to the reference beam, and with better energy overlapping in the target area.

The principle of acoustic lenses has been known for a long time. However, the correction of aberrations by 3D-printed acoustic lenses in the context of transcranial ultrasounds has not been proposed until recently and without taking into account the phenomenon of absorption. The spatial resolution provided by state of art 3D printers allows for a more accurate focusing than that achieved by multi-element arrays. Improvements in data acquisition for lens design, achieved through the use of more precise physics, could be masked by developing multi-element arrays with an effective radiation surface much larger than the simulation voxel section. However, when 3D lenses are used, these improvements are not masked, constituting an open and challenging line of research.

| | | | Point_1 | Point_2 | Point_3 | Point_4 | Point_5 |
|---|---|---|---|---|---|---|---|
| Focus distance | (mm) | | 45.8 | 59.4 | 64.0 | 86.2 | 107.3 |
| Positional deviation (mm) | TRU | $(R, z)$ | 0.105, -0.201 | 0.090,0.6943 | 0.094, 0.080 | 0.319, 4.872 | 0.331, 10.783 |
| | TRA | $(R, z)$ | 0.076, -0.165 | 0.077, 0.466 | 0.074, 0.101 | 0.160, 4.071 | 0.169, 10.863 |
| | Water | | 0.009, 0.103 | 0.031, 0.359 | 0.008,0.047 | 0.043, 1.367 | 0.041, 1.932 |
| Radius of gyration (mm) | TRU | $(k_R, k)$ | 0.476, 1.515 | 0.558, 2.207 | 0.527, 1.552 | 0.972, 4.095 | 1.010, 4.749 |
| | TRA | $(k_R, k)$ | 0.524, 1.608 | 0.554, 2.186 | 0.562, 1.896 | 0.759, 3.310 | 0.938, 4.665 |
| | Water | | 0.398, 1.301 | 0.495, 1.894 | 0.558, 2.351 | 0.678, 3.709 | 0.908, 4.862 |
| Orientation (º) | TRU | $\varphi$ | 2.181 | 1.809 | 4.194 | 3.883 | 3.9107 |
| | TRA | $\varphi$ | 1.873 | 1.644 | 1.934 | 3.208 | 3.3452 |
| | Water | | 0.107 | 0.863 | 0.041 | 0.2175 | 0.164 |
| Volume (mm³) | TRU | $V$ | 7.8589 | 16.693 | 9.809 | 57.807 | 68.8402 |
| | TRA | $V$ | 8.270 | 16.736 | 14.306 | 44.272 | 60.1662 |
| | Water | $V_{ref}$ | 5.239 | 11.410 | 17.47 | 42.336 | 63.6829 |
| Overlapping | TRU | $I(\%)$ | 69.37 | 70.86 | 84.77 | 63.01 | 69.12 |
| | TRA | $I(\%)$ | 66.59 | 72.54 | 79.35 | 74.24 | 81.17 |

Table 1 .- Focusing indicators (-3dB) of the aberrated-corrected scans. Theoretical gold standard emission

| | | | Point_1 | Point_2 | Point_3 | Point_4 | Point_5 |
|---|---|---|---|---|---|---|---|
| Focus distance | (mm) | | 45.8 | 59.4 | 64.0 | 86.2 | 107.3 |
| Positional deviation (mm) | TRU | $(R, z)$ | 0.084, -0.206 | 0.316, 1.675 | 1.415, -0.434 | 0.685, 5.925 | 0.653, 9.622 |
| | TRA | $(R, z)$ | 0.009, -0.008 | 0.155, 1.112 | 1.855, 0.612 | 0.415, 5.800 | 0.547, 9.745 |
| | Water | | 0.004, -0.005 | 0.045, 0.213 | 0.015, 0.143 | 0.054, 1.856 | 0.049, 4.551 |
| Radius of gyration (mm) | TRU | $(k_R, k)$ | 0.679, 2.154 | 1.349, 3.783 | 3.719, 3.939 | 1.417, 5.129 | 1.532, 6.789 |
| | TRA | $(k_R, k)$ | 0.752, 2.374 | 1.031, 3.348 | 4.282, 5.624 | 1.365, 5.092 | 1.581, 6.757 |
| | Water | | 0.533, 1.741 | 0.673, 2.568 | 0.749, 2.827 | 0.925, 4.534 | 1.151, 6.293 |
| Orientation (º) | TRU | $\varphi$ | 2.8438 | 11.078 | 19.0564 | 2.608 | 5.007 |
| | TRA | $\varphi$ | 2.5016 | 6.671 | 46.653 | 1.399 | 5.913 |
| | Water | | 0.0801 | 0.0716 | 0.037 | 1.156 | 0.8326 |
| Volume (mm³) | TRU | $V$ | 23.7515 | 57.341 | 75.594 | 189.953 | 288.163 |
| | TRA | $V$ | 26.9076 | 51.388 | 86.215 | 168.181 | 282.458 |
| | Water | $V_{ref}$ | 14.3789 | 32.934 | 53.717 | 109.515 | 221.652 |
| Overlapping | TRU | $I(\%)$ | 69.15 | 68.72 | 60.84 | 59.82 | 60.76 |
| | TRA | $I(\%)$ | 63.07 | 75.21 | 54.57 | 65.88 | 63.428 |

Table 2 .- Focusing indicators (-6dB) of the aberrated-corrected scans. Theoretical gold standard emission

| Point_4 | | | -3dB | | | -6dB | |
|---|---|---|---|---|---|---|---|
| | | | TRU | TRA | | TRU | TRA |
| Positional deviation (mm) | $(R, z)$ | | 0.928, 7.705 | 0.298, 6.266 | | 1.112, 7.765 | 0.719, 7.755 |
| Radius of gyration (mm) | $(k_R, k)$ | | 1.337, 4.351 | 1.243, 4.102 | | 1.825, 4.768 | 1.983, 4.889 |
| Orientation (º) | $\varphi$ | | 7.597 | 1.4538 | | 3.6300 | 1.1628 |
| Volume (mm³) | $V$ | | 75.95 | 46.6006 | | 230.8927 | 199.7917 |
| Overlapping | $I(\%)$ | | 38.72 | 58.85 | | 22.53 | 26.59 |

Table 3.- Focusing indicators of aberrated-corrected scans, at Point_4. Compete simulation.

**Acknowledgments**

This work was partially supported by the Spanish "Ministerio de Economia y Competitividad" under the project TEC2015-68076-R. The authors would like to thank José Sepúlveda, director of *Asociación I2CV*, for his important input and scientific support.

**Figure captions**

Figure 1. a) Sagital cross sections of the whole computational domain and of the integration domains of configurations Point_1(white) Point_2 (yellow) and Point_3 (orange) b-d) Sagital cross sections of computational domain at experiments Point_1 to Point_3 respectively. e-f) Transversal cross sections of computational domain at experiments Point_4 and Point_5. In figures b to f the internal and external concentric curves represent respectively position of phase pattern and position of transducer, being their separation the maximum thickness of the lens.

Figure 2. Sagital cross section of -6dB focal spots of the aberrated beams at point_2 obtained (a) without any lens, (b) with phase corrected by TRU and (c) by TRA.

Figure 3. Focusing indicators (-3dB) of the reference focal spot and aberrated-phase-corrected focal spots

Figure 4. Details of 3D lens for Point_4: a) Visions of the 3D printing STL file; b) colour map representation of the lens thickness developed respectively by TRA and TRU methods; c) transversal section of computational domain. Colour represents wave speed. Lens and backing (green), water and brain (blue) and skull (variable)

Figure 5. a-b) Sagital cross sections of the stationary energy distribution simulated by ***theoretical gold standard emission*** at configurations Point_1 and Point_3 at the fluid media. c-d) Coronal cross sections at different planes of the stationary energy distribution simulated by ***complete simulation*** at configuration Point_4 at the fluid media (lens and skull modelled as solids).

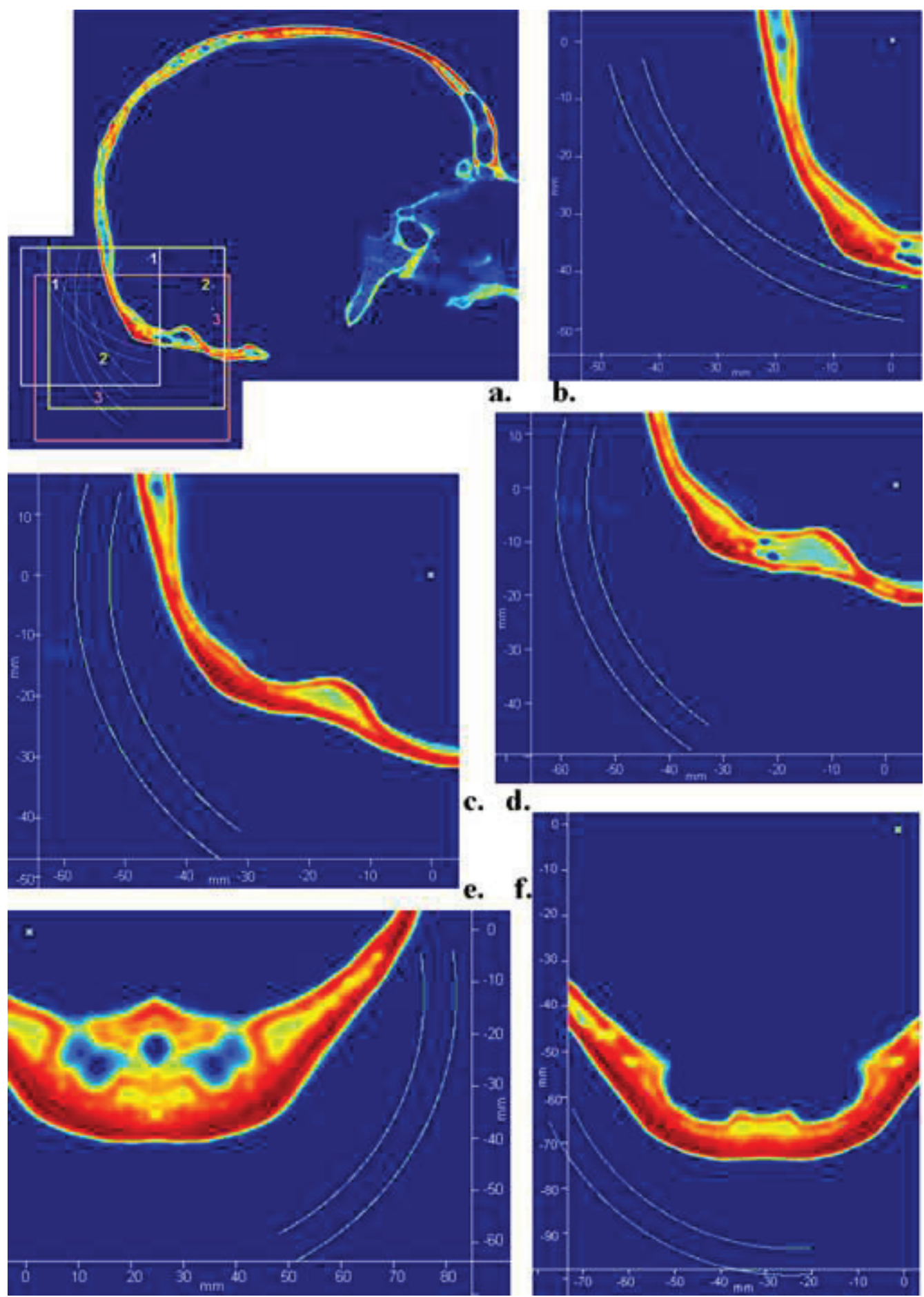

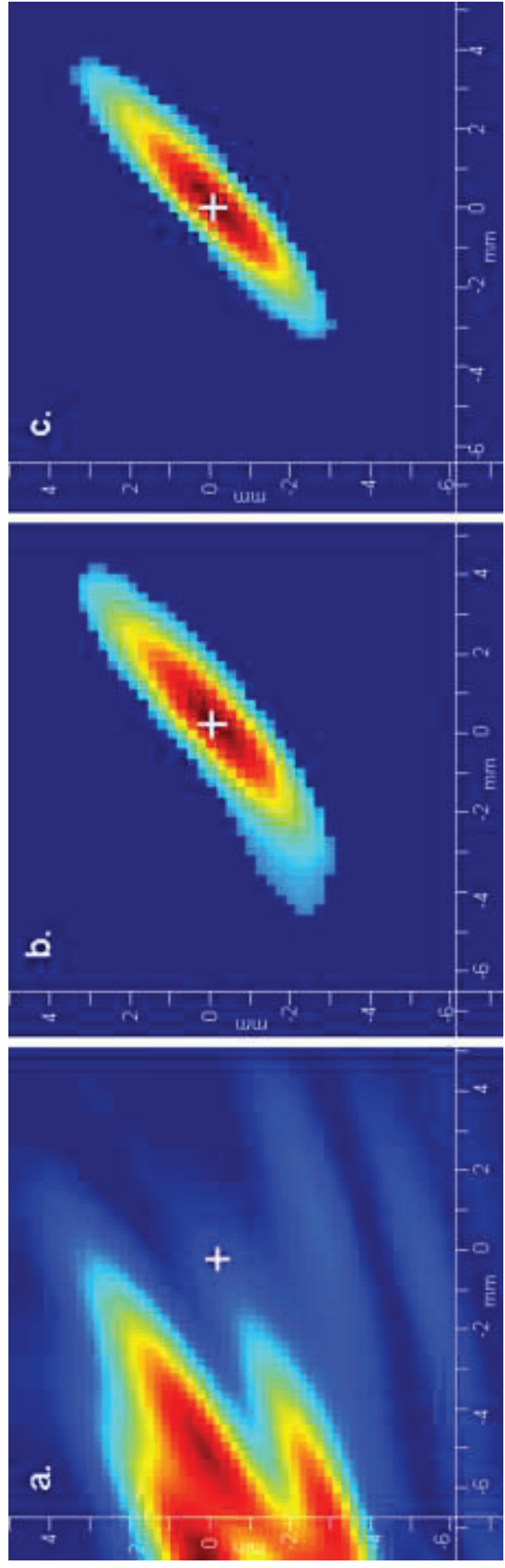

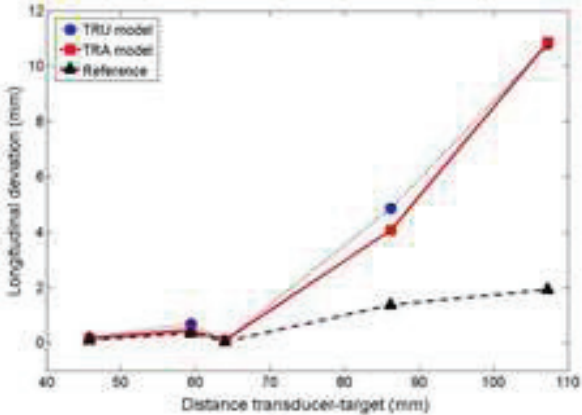

a.

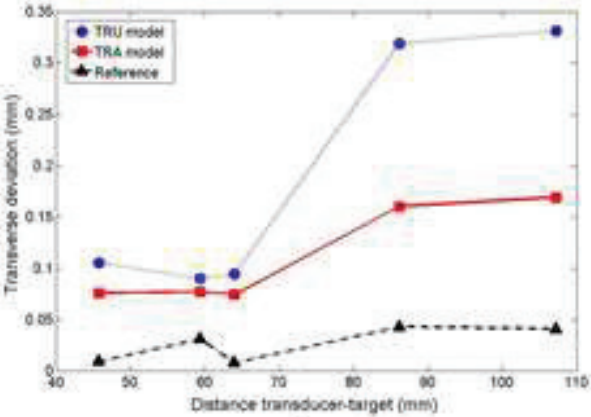

b.

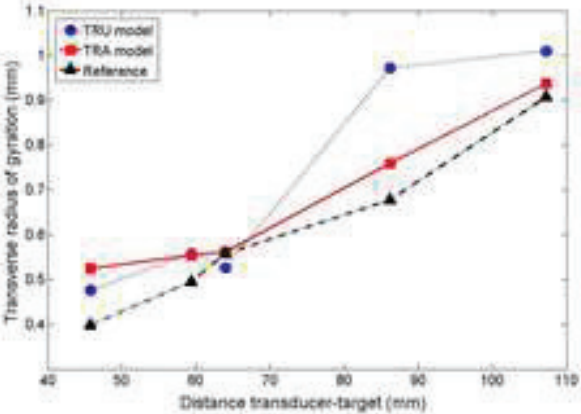

c.

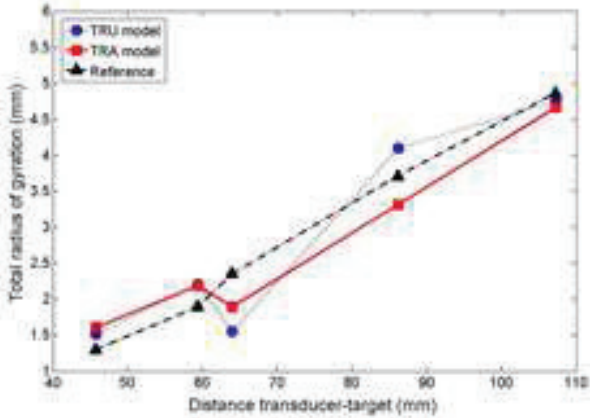

d.

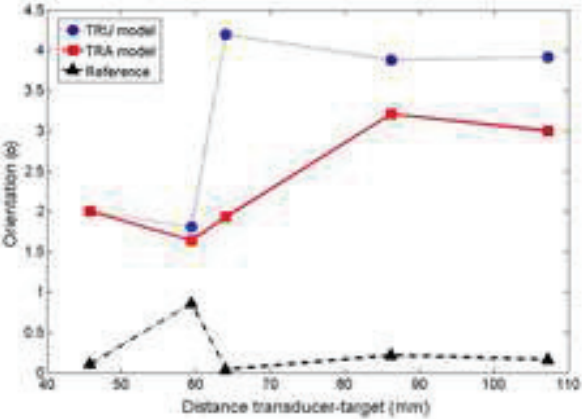

e.

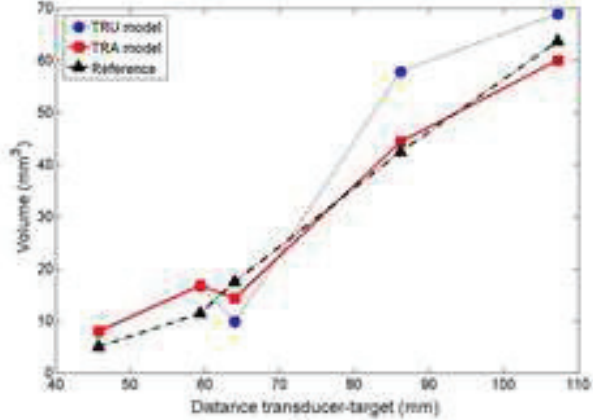

f.

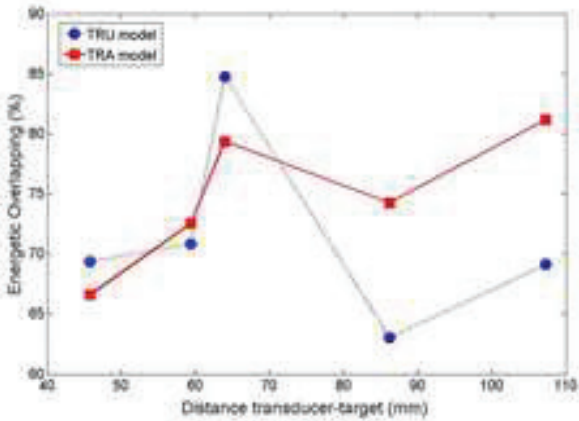

g.

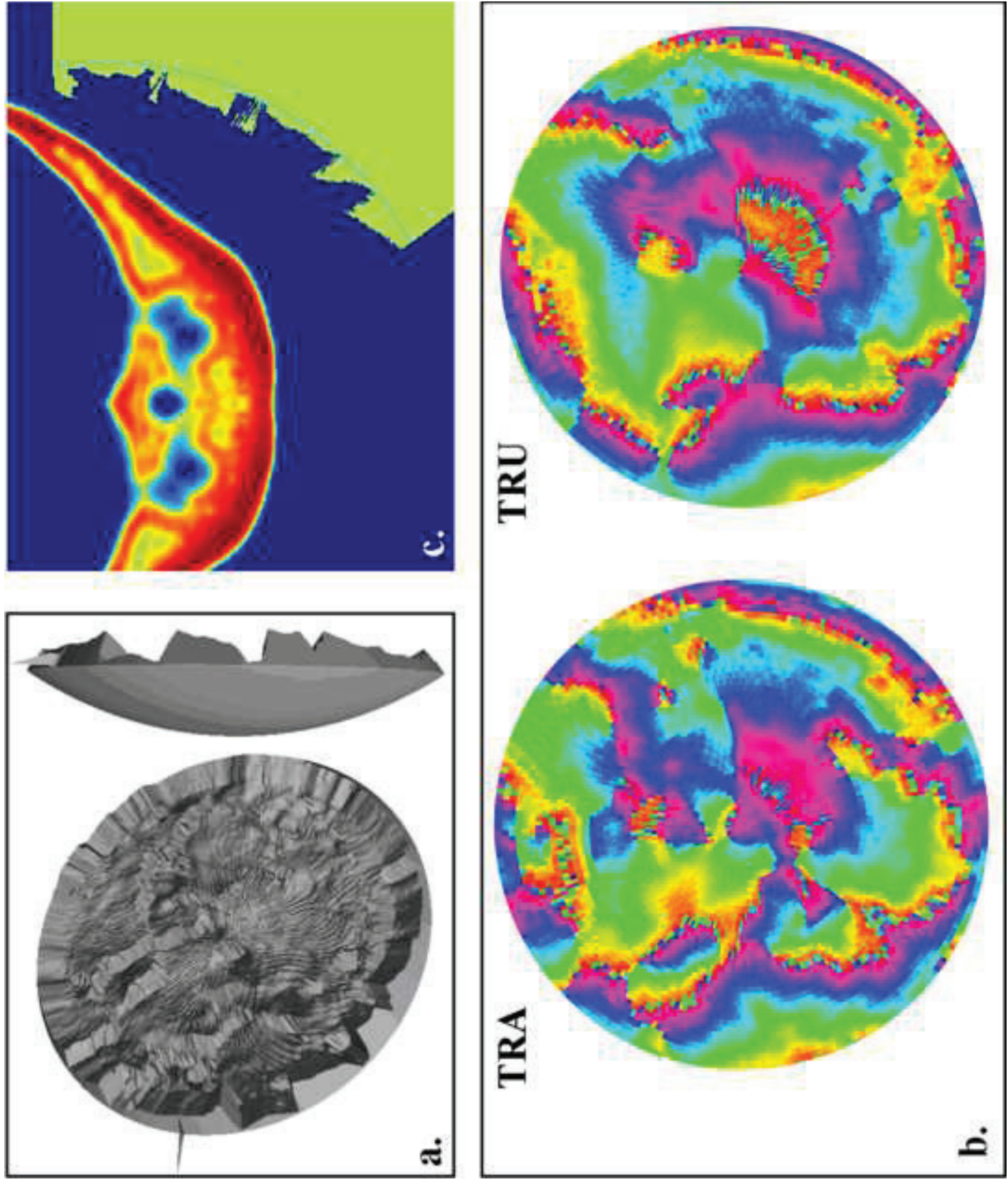

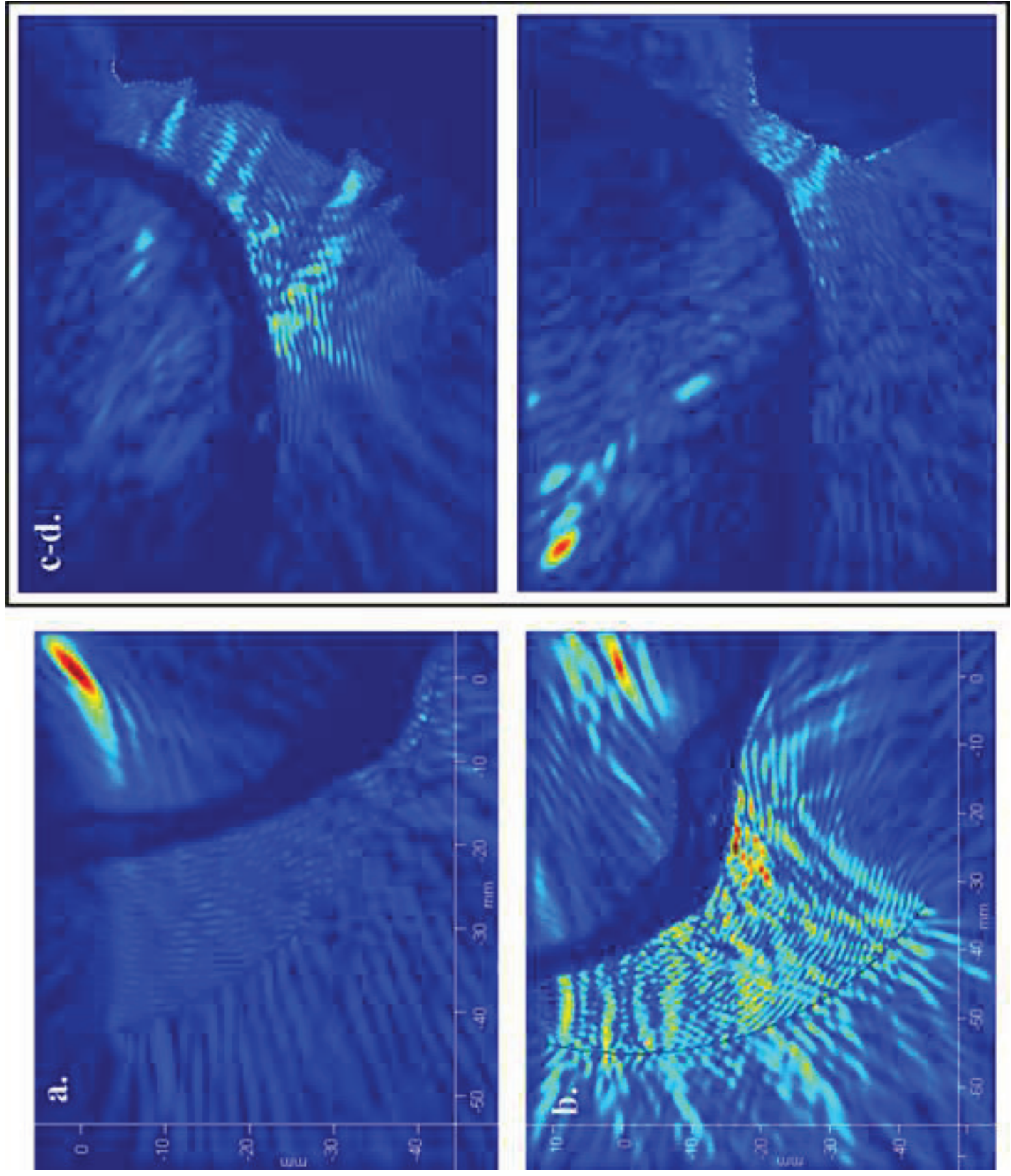